\documentclass[prl,twocolumn,graphicx]{revtex4}
\usepackage{epsfig,amsmath}

\begin{document}

\newcommand{\veck}{\mathbf{k}}
\newcommand{\vecA}{\mathbf{A}}

\newcommand{\vecp}{\mathbf{p}}
\newcommand{\vecr}{\mathbf{r}}

\title{Coulomb repulsion and quantum-classical correspondence in laser-induced
nonsequential double ionization}
\author{C. Figueira de Morisson Faria}
\altaffiliation{Present address: Quantum Optics Group, Institut f\"ur theoretische Physik, Universit\"at Hannover, Appelstr. 2, 30167 Hannover, Germany}
\affiliation{Max-Born-Institut, 
Max-Born-Str. 2A, 12489 Berlin, Germany}
\author{X. Liu}
\affiliation{Max-Born-Institut, 
Max-Born-Str. 2A, 12489 Berlin, Germany}
\author{H. Schomerus}
\affiliation{Max-Planck-Institut f\"{u}r Physik komplexer Systeme, 
N\"{o}thnitzer Str. 38, 01187 Dresden, Germany}
\author{W. Becker}
\affiliation{Max-Born-Institut, 
Max-Born-Str. 2A, 12489 Berlin, Germany}
\date{\today}

\begin{abstract}
The influence of electron-electron Coulomb repulsion on
nonsequential double ionization of rare-gas atoms is investigated.
Several variants of  the quantum-mechanical transition amplitude are evaluated that differ by the form of the inelastic electron-ion rescattering and whether
or not  Coulomb
repulsion between the two electrons in the final state is included.
For high laser intensity, an entirely classical model is formulated that simulates the rescattering scenario. 
\end{abstract}
\maketitle

Multiple ionization of atoms by intense laser fields can proceed via different quantum-mechanical pathways. 
The atom or ion may be ionized step by step such that the transition amplitude is the product of the amplitudes for single ionization. 
If such a factorization is not possible, one speaks of \textit{nonsequential} multiple ionization. 
Various physical mechanisms may be envisioned to underlie the latter, but electron-electron correlation is a necessary precondition. 

The actual presence of the nonsequential double ionization (NSDI) pathway was inferred long ago \cite{l'hui83} from data at 1053 nm at rather low intensity, 
but the  mechanism responsible for it could only be identified after the reaction-microscope technique  provided much more detailed information about the process than was available before \cite{basicrefs}. 
In principle, this technique is capable  of analyzing double ionization in terms of all six momentum components of the  
ion and one electron, while earlier experiments were only able to yield \textit{total} double-ionization rates. 
To the extent that the photon momentum can be neglected,  
this is synonymous with a complete kinematical characterization of the process. As a result, rescattering has emerged as the dominant mechanism, as it is for  high-order harmonic generation and high-order above-threshold ionization \cite{corkum}. 
For a review of recent developments, see Ref.~\cite{advan}.

An exact description of NSDI is, in practice,  only possible for helium and requires the solution of the six-dimensional time-dependent Schr\"odinger equation \cite{taylor,muller}.  
Various approximations and models have been considered,  
 in particular, one spatial dimension for each electron \cite{eberly,lein}, or neglecting the backaction of the outer electron on the inner 
 \cite{watson}, or density-functional methods \cite{bauer}. Classical-trajectory calculations \cite{chen} 
yield 
reasonable agreement with the data 
far above the threshold. The exact quantum-mechanical transition amplitude has been analyzed in 
terms of Feynman diagrams  \cite{faisal}.
The basic diagram that contains rescattering has been evaluated by several groups \cite{fb,KBRS,gorpop}.
It incorporates (i) tunneling of the first electron and (ii) its subsequent motion in the laser field, (iii) the 
inelastic-rescattering collision with the second (up to this time bound) electron, 
and (iv) the propagation of the final two electrons in the laser field. It does not contain, for example, the interaction of the electron in the intermediate state with the ion nor the interaction 
of the final electrons with the ion or with each other.

In this Letter, we incorporate the Coulomb repulsion between the two electrons in the final state into
the basic quantum-mechanical transition amplitude of NSDI,
and compare with experiments.
Moreover, we formulate a classical inelastic-rescattering model that reproduces the quantum-mechanical results for high laser intensity.

\begin{figure}
\begin{center}
\includegraphics[width=6cm]{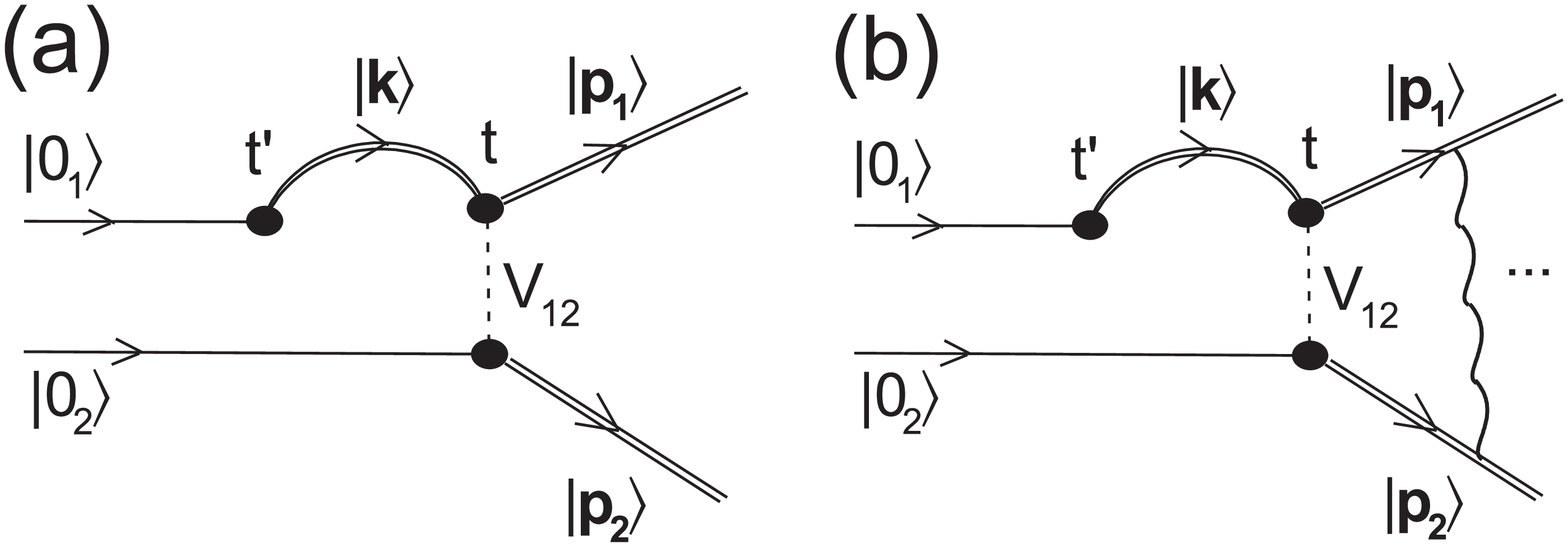}
\end{center}
\caption{Feynman diagrams corresponding to the transition amplitude (\ref{rescatt}), (a) without and (b) with electron-electron repulsion between the two electrons in the final state. 
The vertical wavy line and the dots in (b) indicate Coulomb interaction, which is exactly accounted for by the two-electron Volkov solution. 
The dashed vertical line represents the electron-electron interaction $V_{12}$ by which the second electron is set free.}\label{fig:1}
\end{figure}

The quantum-mechanical transition amplitude formalizing the assumptions of the rescattering model 
is \cite{GPKB02}
\begin{eqnarray}
M&=&-\int_{-\infty }^{\infty }dt\int_{-\infty }^{t}dt^{\prime }\langle 
\psi _{\mathbf{p}_{1}\mathbf{p}_{2}}^{\mathrm{(V)}}(t)| V_{12}
\nonumber\\
&&\times U_{1}^{\mathrm{(V)}}(t,t^{\prime })V_1|\psi _{0}^{(1)}(t^{\prime })\rangle 
\otimes |\psi_0^{(2)}(t)\rangle.  \label{rescatt}
\end{eqnarray}
It is pictorialized in the Feynman diagrams of Fig.~\ref{fig:1}:
Initially, both electrons are bound in their (uncorrelated) ground states   
$| \psi _{0}^{(n)}(t')
\rangle =e^{i|E_{0n}|t^{\prime }}| \psi _{0}^{(n)}\rangle$, 
where $E_{0n}$  is the binding energy of the $n$th electron.
At the time $t^{\prime }$, the first electron is released from the binding potential $V_1$ through
tunneling ionization, whereas the second electron remains bound.
Subsequently, the first electron propagates in the continuum
described by the Volkov time-evolution operator $U_1^{\mathrm{(V)}}(t,t^{\prime })$, gaining energy from the field. 
At the later time $t$, it dislodges  the second electron in an inelastic collision mediated by 
the interaction $V_{12}$, which is accounted for in the lowest-order Born approximation.
Throughout the paper, we compare two possible choices for this 
interaction: the Coulomb interaction $V_{12}\sim |\mathbf{r}_1 - \mathbf{r}_2|^{-1}$ 
and a three-body contact interaction $V_{12}\sim \delta(\mathbf{r}_1 - \mathbf{r}_2)\delta(\mathbf{r}_1)$. 
The latter might be interpreted as an \textit{effective} electron-electron interaction 
on the background of the ion.  
For the final two-electron state   
$|\psi _{\mathbf{p}_{1}\mathbf{p}_{2}}^{\mathrm{(V)}}(t)\rangle$
 with asymptotic momenta 
$\mathbf{p}_{1}$, $\mathbf{p}_{2}$, we take  
the correlated (outgoing)
two-electron Volkov state \cite{bf94} 
\begin{eqnarray}
&&|\psi^{\mathrm{(V)}}_{\vecp_1\vecp_2}(t)\rangle=|\psi^{\mathrm{(V)}}_{\vecp_1}(t) \rangle \otimes  |\psi^{\mathrm{(V)}}_{\vecp_2}(t)\rangle\nonumber\\
&&\times _1F_1(-i\gamma,1;i(|\vecp||\vecr|-\vecp\cdot\vecr))e^{-\pi\gamma/2}\Gamma(1+i\gamma),\qquad\label{2v}
\end{eqnarray}
which exactly accounts for their Coulomb repulsion [Fig.~\ref{fig:1}(b)], and compare it with the product state of one-electron
Volkov states $|\psi^{\mathrm{(V)}}_{\vecp_i}(t)\rangle$ [Fig.~\ref{fig:1}(a)].  Here $\vecp=(\vecp_1-\vecp_2)/2$, $\vecr=\vecr_1-\vecr_2$, and $\gamma=1/(2|\vecp|)$ 
(Coulomb repulsion is turned off by setting $\gamma=0$).  

In order to evaluate the multiple integrals in the transition amplitude (\ref{rescatt}), we expand the Volkov propagator  
$U^{\mathrm{(V)}}_1(t,t')=\int d^3 \veck |\psi_\veck^{\mathrm{(V)}}(t)\rangle \langle \psi_\veck^{\mathrm{(V)}}(t')|$ in terms of the Volkov states,
$\langle \vecr |\psi_\veck^{\mathrm{(V)}}(t)\rangle=(2\pi)^{-3/2} \exp\{i[\veck + \vecA(t)]\cdot\vecr\}$ $\exp[iS_\veck (t)]$, with $S_\veck(t)= -(1/2)\int^t d\tau [\veck + \vecA(\tau)]^2$ 
the action of a free electron in the presence of the laser field described by the vector potential $\vecA(t)$. Since the Volkov solutions are plane waves (with time-dependent momentum), 
the spatial integrals yield the two form factors
\begin{eqnarray}
&& V_{\mathbf{pk}}=\langle \mathbf{p}_{2}+\mathbf{A}(t),\mathbf{p}_{1}+
\mathbf{A}(t)| V_{12}| \mathbf{k}+\mathbf{A}(t),\psi_{0}^{(2)}\rangle
,\quad 
\label{v12}\\
&& V_{\mathbf{k}0}=\langle \mathbf{k}+\mathbf{A}(t^{\prime })|
V| \psi _{0}^{(1)}\rangle ,\label{v10}
\end{eqnarray}
which, for the contact and Coulomb potentials that we consider, can be obtained in closed form. 
The remaining integrals over the intermediate-state momentum $\veck$, the ionization time $t'$, 
and the rescattering time $t$ are evaluated 
by saddle-point integration, 
 which is justified for the high ponderomotive energies $U_P$ of the experiments.

The standard saddle-point approximation reduces the five-dimensional
integration over $\veck, t$, and $t'$ to a sum over the complex solutions $\veck_s,t_s,t'_s\ (s=1,2,\dots)$ of the saddle-point equations.
 The transition amplitude
 \begin{equation}
 M^{(\mathrm{SPA})} =\sum_{s}\frac{(2\pi i)^{5/2}V_{\mathbf{p}\mathbf{k}_{s}}V_{\mathbf{k}_{s}0}
 }{\sqrt{\det S_{\mathbf{p}}^{\prime \prime }(t,t^{\prime },\mathbf{k})|_{s}}}
 e^{iS_{\mathbf{p}}(t_{s},t_{s}^{\prime },\mathbf{k}_{s})}
 \label{sadresc}
 \end{equation}
 is the coherent superposition of the contributions of all relevant saddle points. Here, the various exponentials in the amplitude (\ref{rescatt}) have been collected into the action $S_\vecp(t,t',\veck)$.
The method is explained in detail in Refs.~\cite{GPKB02,nsdiuni}. 
In rescattering problems, 
the complex solutions come in pairs that approach each other very closely near the classical cutoffs. In this case, 
the standard saddle-point approximation becomes inapplicable. We here employ a
so-called uniform approximation \cite{atiuni,nsdiuni},
which invokes the same information
on the saddles but works regardless of their separation.

Both in the saddle-point and in the uniform approximation, the upshot of employing the correlated two-electron Volkov state (\ref{2v}) is very simple: the result of the uncorrelated two-electron Volkov state just has to be augmented by the factor  
$2\pi\gamma/[\exp(2\pi\gamma)-1]$ \cite{footnote}.

The consequences are illustrated in
the next figures, 
where $V_{12}$ is the contact interaction or the Coulomb interaction, respectively, and the parameters correspond to the data for argon \cite{resolvedar2,resolvedar1} (note \cite{nevsar}) and neon \cite{neon}.
We follow the presentation of the experimental data and decompose 
the final-state momenta into their components parallel and perpendicular to the (linearly polarized) laser field, so that $\vecp_i \equiv (p_{i\parallel},\vecp_{i\perp})\ (i=1,2)$. 
Then, 
we present density plots of the double-ionization probability as a function of the parallel components $p_{1\parallel}$ and $p_{2\parallel}$, 
while the transverse components $\vecp_{i\perp}$ are partially or entirely integrated over.

\begin{figure}
\begin{center}
\includegraphics[width=8cm]{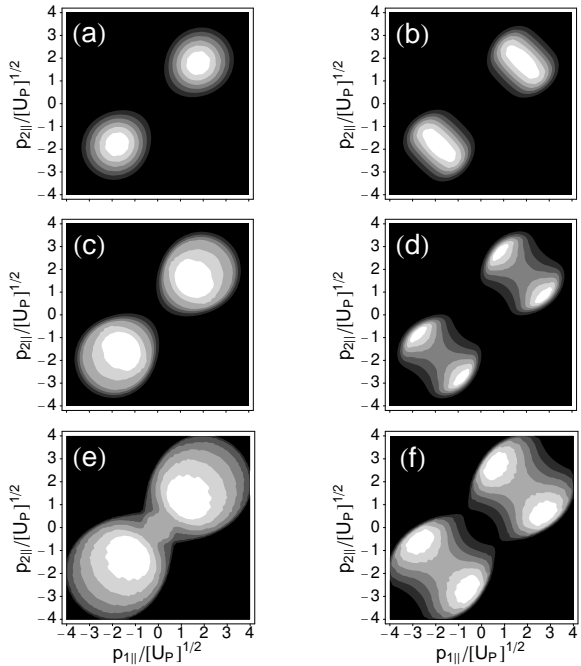}
\end{center}
\caption{
Comparison of the double-ionization probability densities
without (left-hand column) and with (right-hand column)
electron-electron repulsion in the final state,
as a function of the electron momenta parallel to the laser field.
The  interaction $V_{12}$ is specified by the three-body contact interaction.
Parameters are for argon ($E_{01}=0.58$ a.u., $E_{02}=1.015$ a.u.),
the laser frequency is $\omega=0.057$ a.u. (Ti:Sa). 
Panels (a) and (b): $I=2.5 \times 10^{14}$ Wcm$^{-2}$ ($U_P=0.54$ a.u),
$|\mathbf{p}_{1\perp}|\ge 0.5$ a.u.; Ref.~\cite{resolvedar2};
(c) and (d): as before, but  with $|\mathbf{p}_{1\perp}|\le 0.5$ a.u.;
(e) and (f): 
$I= 4.7\times10^{14}$ Wcm$^{-2}$ ($U_P=1.0$ a.u),
$|\mathbf{p}_{1\perp}|$ or $|\mathbf{p}_{2\perp}|$
$\le 0.1$ a.u., Ref.~\cite{resolvedar1}. }
\label{fig:repulcon}
\end{figure}

\begin{figure}
\begin{center}
\includegraphics[width=8cm]{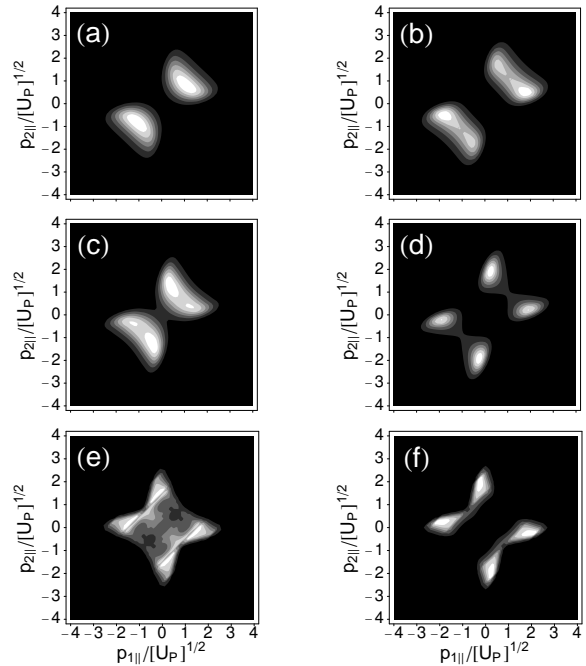}
\end{center}
\caption{Same as Fig.~\ref{fig:repulcon}, but with $V_{12}$ specified by the Coulomb interaction.}
\label{fig:repulcou}
\end{figure}

In general,
final states where the two electrons have similar (vector) momenta are
suppressed by their Coulomb repulsion, since the two electrons are set free simultaneously by the interaction $V_{12}$.
In the $(p_{1\parallel},p_{2\parallel})$-correlation plots, this tends to reduce the population on the diagonal.
The effect is most pronounced if both transverse momenta are small, and still sizable if only one is small [Figs.~\ref{fig:repulcon}, \ref{fig:repulcou} (c-f)]. 
When one transverse momentum is large,
the effect is only moderate [Figs.~\ref{fig:repulcon}, \ref{fig:repulcou} (a,b)].

Figures \ref{fig:repulcon}, \ref{fig:repulcou} (a-d) are for the parameters
of Ref.~\cite{resolvedar2}. 
While inconclusive if one of the transverse momenta is large
[Figs.~\ref{fig:repulcon}, \ref{fig:repulcou} (a,b)],
including the Coulomb repulsion leads to improved agreement if one of them is small
[Figs.~\ref{fig:repulcon}, \ref{fig:repulcou} (c,d)], but
the resolution of the experiment does not allow one to settle
either on Fig.~\ref{fig:repulcon}(d) or on Figs.~\ref{fig:repulcou}(c,d). 
Closer scrutiny of 
Figs.~\ref{fig:repulcou}(a)-(d) unveils a lack of symmetry with respect to the diagonal, which, possibly, has been observed in Ref.~\cite{resolvedar2}. It arises since the
Coulomb form factor (\ref{v12}) is only symmetric under interchange of
{\em all} momentum components $\vecp_1 \leftrightarrow \vecp_2$
[in the final state (\ref{2v}), transverse and parallel momentum components
can be interchanged independently].
Higher intensity and more restricted transverse momenta are
investigated in Ref.~\cite{resolvedar1}, and for  very small transverse momenta [Figs.~\ref{fig:repulcon}, \ref{fig:repulcou}(e,f)] we do find 
the characteristic Coulomb pattern of Fig.~\ref{fig:repulcou} reflected in the experiment.
The data agree better with Fig.~\ref{fig:repulcou}(e), which does not incorporate final-state repulsion, than with  Fig.~\ref{fig:repulcou}(f), which does.

\begin{figure}[b]
\begin{center}
\includegraphics[width=8cm]{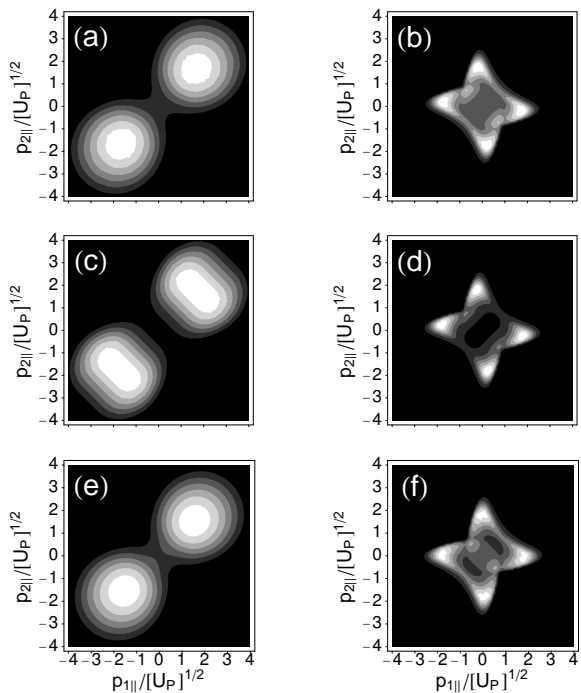}
\end{center}
\caption{Double-ionization probability densities for neon ($E_{01}=0.9$
a.u., $E_{02}=1.51$ a.u.),
$I= 10^{15}$ Wcm$^{-2}$ ($U_P=2.2$ a.u.),
integrated over all transverse momenta \cite{neon}.
Left panels: $V_{12}$ the contact interaction,
right panels:  $V_{12}$ the Coulomb interaction. The upper four 
are computed from the quantum-mechanical amplitude (\ref{rescatt}),
in the absence (a,b) and presence (c,d) of final-state
 Coulomb repulsion, respectively. The remaining panels (e,f)
are calculated from 
its classical analog (\ref{class}) without final-state
electron-electron repulsion.
}
\label{fig:clqm}
\end{figure}

In neon \cite{neon}, the $\vecp_{i\perp}$-integrated momentum correlation
is quite well reproduced by the contact interaction \cite{GPKB02}. Corresponding results without and with final-state
Coulomb repulsion are shown in  Figs.\ \ref{fig:clqm}(a,b) and \ref{fig:clqm}(c,d), respectively. We
find that taking into account the final-state Coulomb repulsion \textit{does not} improve the agreement.
The very simplest model -- a  contact interaction and no final-state repulsion -- works best. This suggests that the $V_{12}$ contact interaction
can be interpreted as a reasonable zeroth-order \textit{effective} interaction that takes into account the presence of the ion, which shields the long-range interaction between the returning and the bound electron.

The rescattering diagrams of Fig.~\ref{fig:1}, though fully quantum-mechanical, stimulate a classical interpretation. In what follows, we recast their physical content into an entirely classical expression. First, let us enumerate the quantum features inherent  in the transition amplitude (\ref{rescatt}): (i) the electron enters the field via tunneling, 
which enforces, in the saddle-point solution,  complex $t'_s$,
$t_s$, and $\veck_s$; (ii)  the contributions of the individual saddle points 
are added coherently in the sum (\ref{sadresc}); (iii) in principle, the atom
can absorb an arbitrary number of photons from the laser field, that is, 
double ionization still occurs, though at a much reduced rate, if the maximal
kinetic energy of the returning electron is less than  $|E_{02}|$; (iv) the
electron wave packet spreads during its propagation in the continuum.
However, we do not expect the quantum features (ii) and (iii) to have a significant impact on the yields of double ionization far above the threshold, and feature (iv) is only relevant when  long orbits contribute \cite{PRL02}.

In order to check this surmise, we consider the classical model
\begin{eqnarray}
|M_{\mathrm{class}}|^2 \sim  \int dt' R(t') \delta \left(E_{\rm ret}(t)-\frac{(\vecp_1+\vecA(t))^2}{2m} \right.\nonumber\\
\left.- \frac{(\vecp_2+\vecA(t))^2}{2m} - |E_{02}|\right) |V_{\vecp \veck(t)}|^2.\label{class}
\end{eqnarray}
Here, the electron appears in the continuum with zero velocity at the time $t'$ at the time-dependent rate $R(t')$, which describes the quantum-mechanical tunneling process
 (cf. the above feature (i);
we will use a simple tunneling rate \cite{ll}). For each ionization time $t'$, the return times $t\equiv t(t')$, and the corresponding kinetic energies $E_\mathrm{ret}(t)$ and  drift momenta $\veck(t)$ are calculated
along the lines of
the classical simple-man model \cite{corkum}. They correspond to the saddle-point solutions $t_s$, except that they are real \cite{clfoot}. The $\delta$-function in Eq.~(\ref{class}) 
expresses energy conservation in the inelastic collision that sets free the second electron.   
The actual distribution of the final momenta is governed by the form factor $|V_{\vecp \veck(t)}|^2$. In Eq.~(\ref{class}), the \textit{probabilities} of the various orbits $s=0,1,\dots$ are added so that their 
contributions cannot interfere, in contrast to the quantum-mechanical amplitude (\ref{sadresc}).
In Fig.\  \ref{fig:clqm}(e) and (f), we present the result of the classical model when the transverse momentum components $\vecp_{i\perp}$ are completely integrated over.
They agree with the quantum-mechanical results of Figs.~\ref{fig:clqm}(a) and \ref{fig:clqm}(b) quite well. 
A similar conjecture was derived from the comparison of one-dimensional classical-trajectory and quantum calculations  \cite{eberly}. 
However, the quantum distributions are slightly wider and the low-density regions are 
enhanced in comparison with the classical distributions, reflecting the fact  that the quantum distributions extend into the classically forbidden region.

In summary, we have investigated the effects of electron-electron repulsion in the final two-electron state of nonsequential double ionization. The calculations allow us to conclude that footprints of electron-electron repulsion have been observed in experiments where the transverse momentum of one of the electrons is small, with better agreement when the repulsion occurs during the
rescattering process rather than in the final state of the ionized electrons.
We suggest that experiments in which {\em both} electrons are
restricted to small transverse momenta would be most incisive. 
If the laser intensity 
is high enough, the quantum-mechanical  momentum distributions can be well reproduced in a purely classical model.

Added note: Very recently, a paper by M. Weckenbrock \textit{et al.}, Phys. Rev. Lett. \textbf{91}, 123004 (2003), was published, which addresses similar questions.

We benefitted from discussions with A. Becker, E. Eremina, S.P. Goreslavski, D.B. Milo\v{s}evi\'c, S.V. Popruzhenko, H. Rottke, and W. Sandner, and we are greatly indebted to E. Lenz for help with the code. This work was supported in part by the Deutsche Forschungsgemeinschaft.


\begin{thebibliography}{99}
\bibitem{l'hui83}A. l'Huillier, L.A. Lompre,  G. Mainfray, and C. Manus, Phys. Rev. A \textbf{27}, 2503 (1983).
\bibitem{basicrefs}  T. Weber \textit{et al.}, Phys. Rev. Lett. \textbf{ 84}, 443 (2000); R. Moshammer \textit{et al.},  Phys. Rev. Lett. \textbf{ 84}, 447 (2000).  
\bibitem{corkum}  P.B. Corkum, Phys. Rev. Lett. \textbf{71}, 1994 (1993).
\bibitem{advan}R. D\"orner \textit{et al.}, Adv. At., Mol., Opt. Phys. \textbf{48}, 1 (2002).
\bibitem{taylor}D. Dundas, K.T. Taylor, J.S. 
Parker, and E. S. Smyth, J. Phys. B \textbf{ 32}, L231 (1999).
\bibitem{muller} H. G. Muller, Opt. Express \textbf{8}, 425 (2001).
\bibitem{eberly}R. Panfili, S.L. Haan, and J.H. Eberly, Phys. Rev. Lett. \textbf{89}, 113001 (2002); S.L. Haan, P.S. Wheeler, R. Panfili, and J.H. Eberly, Phys. Rev. A \textbf{66}, 061402(R) (2002). 

\bibitem{lein}M. Lein, E.K.U. Gross, and V. Engel, Phys. Rev. 
Lett. \textbf{ 85}, 4707 (2000).
\bibitem{watson}  J.B. Watson \textit{et al.}, Phys. Rev. Lett. \textbf{78}, 1884 (1997).
\bibitem{bauer} D. Bauer and F. Ceccherini, Opt. Express \textbf{8}, 377 (2001).
\bibitem{chen}L.-B. Fu, J. Liu, and S.-G. Chen, Phys. Rev. A \textbf{65}, 021406(R) (2002).
\bibitem{faisal}A. Becker and F.H.M. Faisal, J. Phys. B \textbf{29}, L197 (1996).
\bibitem{fb} A. Becker and F.H.M. Faisal, Phys. Rev. Lett. \textbf{84}, 3546 (2000); \textit{ibid.} \textbf{89}, 193003 (2002).
\bibitem{KBRS}R. Kopold, W. Becker, H. Rottke, and W. Sandner, Phys. Rev. Lett. \textbf{85}, 3781 (2000).

\bibitem{gorpop} S.V. Popruzhenko and S.P. Goreslavskii, J. Phys. B \textbf{34}, L239 (2001).
\bibitem{GPKB02} S.P. Goreslavskii, S.V. Popruzhenko, R. Kopold, and 
W. Becker, Phys. Rev. A \textbf{ 64}, 053402 (2001).
\bibitem{bf94} A. Becker and F.H.M. Faisal, Phys.\ Rev. A \textbf{ 50}, 
3256 (1994).
\bibitem{nsdiuni}  C. Figueira de Morisson Faria and W. Becker, Laser 
Phys. \textbf{13}, 1196 (2003).

\bibitem{atiuni}  C. Figueira de Morisson Faria, H. Schomerus, and W. Becker,
Phys. Rev. A \textbf{ 66}, 043413 (2002).


\bibitem{footnote} 
This is an excellent approximation and exactly true in so much as in the saddle-point integration 
the intermediate momentum $\veck$ can be treated as real; 
for details, see C. Figueira de Morisson Faria \textit{et al.} (to be published).

\bibitem{resolvedar2}  R. Moshammer \textit{et al.},  
Phys. Rev.\ A \textbf{ 65}, 035401 (2002).
\bibitem{resolvedar1}M. Weckenbrock \textit{et al.},
J. Phys. B \textbf{ 34}, L449 (2001).


\bibitem{nevsar}
In argon, as opposed to neon, there is experimental evidence of an additional ionization mechanism, possibly excitation of the second electron followed by tunneling [B. Feuerstein \textit{et al.}, Phys. Rev. Lett. \textbf{87}, 043003 (2001)], which generates momenta around $p_{1\parallel}=p_{2\parallel}=0$. These are absent in our calculations, which only include \textit{instantaneous} inelastic rescattering. The different behavior of neon and argon has been traced back to the respective ionization and excitation cross sections by V.L. Bastos de Jesus \textit{et al.} (in print).

\bibitem{neon} R. Moshammer \textit{et al.}, J. Phys. B \textbf{36}, L113 (2003). 
 
 \bibitem{PRL02}  Spreading dampens the contributions of the longer orbits, whose influence on the yields is minor, except at channel closings;  see 
 S.V. Popruzhenko, P.A. Korneev, S.P. Goreslavski, and W. Becker, Phys. Rev. Lett. \textbf{89}, 023001 (2002). Only
 the two shortest orbits are included in our calculations.
\bibitem{ll} L.D. Landau and E. M. Lifshitz, {\it Quantum Mechanics} 
(Pergamon,  Oxford, 1977). 

\bibitem{clfoot} In fact, the classical simple-man model for NSDI has been derived from the amplitude (\ref{rescatt}) in Ref.\ \cite{gorpop}.


\end{thebibliography}
\end{document}